# Pt-wedge squeegee cleaning of two-dimensional materials and heterostructures

Emine Yegin[1,2], Doruk Pehlivanoğlu[1,2], T. Serkan Kasırga[2,3*]

[1] Institute of Materials Science and Nanotechnology, Bilkent University, Ankara 06800 Türkiye
[2] Bilkent National Nanotechnology Research Center – UNAM, Ankara 06800 Türkiye
[3] Department of Physics Middle East Technical University, Ankara 06800 Türkiye
[*] Corresponding author: kasirga@bilkent.edu.tr

**Abstract**

The surface of ultra-thin materials plays a crucial role in determining the properties. This is particularly important in two-dimensional (2D) materials where the surface-bulk distinction is no longer present. While mechanical cleaning of two-dimensional materials to remove interfacial and surface contaminants is used to achieve better sample quality, low throughput and the challenging optimization of cleaning procedures hinder their widespread adoption. Here, we report on atomic force microscope (AFM)-based mechanical cleaning with modified AFM cantilevers for high-throughput and easy-to-implement cleaning of 2D materials and their heterostructures. A Pt-wedge is deposited via focused ion beam on the cantilever to improve the mechanical cleaning of samples and streamline the cleaning procedures. We demonstrate that a cleaning rate of 3 µm²/s can be achieved with our modified cantilevers, compared to the 0.01 µm²/s effective cleaning rate in pointy-tip cleaning. As showcases, we demonstrate that monolayer $WS_2$ on h-BN exhibits much sharper photoluminescence (PL) emission at room temperature after AFM cleaning, and $WS_2$ monolayers exhibit a higher quality contacts to cleaned Au electrodes as compared to uncleaned electrodes. We also showed that h-BN encapsulated heterostructures can be cleaned rapidly using the improved method. Overall, our results exhibit a feasible and facile path for the large-scale application of AFM-based cleaning of integrated 2D materials.

**Main Text**

The properties of two-dimensional (2D) materials may vastly differ from their bulk counterpart due to the strong confinement of the electronic degrees of freedom. However, such strong confinement causes proximity effects to play a substantial role in the properties of two-dimensional materials. Regardless of whether the material is synthesized via chemical vapor deposition or mechanically exfoliated from a bulk crystal, a substrate is needed to support the 2D material. The supporting substrate creates an interface with the 2D material and modifies the properties of the materials via the proximity effect, often with spatial non-uniformity[1–7]. Even when 2D materials are transferred over atomically flat and smooth substrates, surface contaminants trapped between the 2D layer and the substrate can result in spatial variation in the dielectric constant, induce strain, and create locally doped or electrostatically gated regions. Such alterations decrease the device performance of 2D materials and also hinder their intrinsic properties.

To mitigate the aforementioned problems, thermal annealing is commonly used in sample fabrication. However, despite improving the overall interface and device quality, thermal annealing still causes localized blisters to form. This results in spatial variation in the sample properties. Performing the transfer in a controlled environment via dry transfer methods also improves the interface quality, yet reproducibility and uniformity at the micron level are still not at the desirable level. Further cleaning with an atomic force microscope (AFM) tip has been performed on graphene[8–11] and recently on other materials[12] for higher quality devices of 2D materials. Interfacial contaminants are pushed to the sides of the sample via scanning with an AFM tip in the contact mode[13]. As a result, superior interface qualities and device performances have been demonstrated.

Despite the promising results, AFM tip cleaning requires stringent optimization of AFM parameters[14–16]. The main challenge is that multiple hours, often overnight, of scanning are required to clean tens of square micrometers of 2D material interfaces. Moreover, when large forces are applied by the tip, the 2D material or the heterostructure can be easily ripped by the large pressure of the atomically pointy AFM tip. Even with typical parameters such as cleaning force, tip radius, and cantilever spring constant, large sample-to-sample variations pose challenges to the reproducible application of AFM tip cleaning on samples. Using a nanospherical AFM tip has been shown to be promising[17] but it possesses similar problems to atomically pointy AFM tips. Despite parallel cantilever scanning strategies having been proposed, they haven't been demonstrated yet.

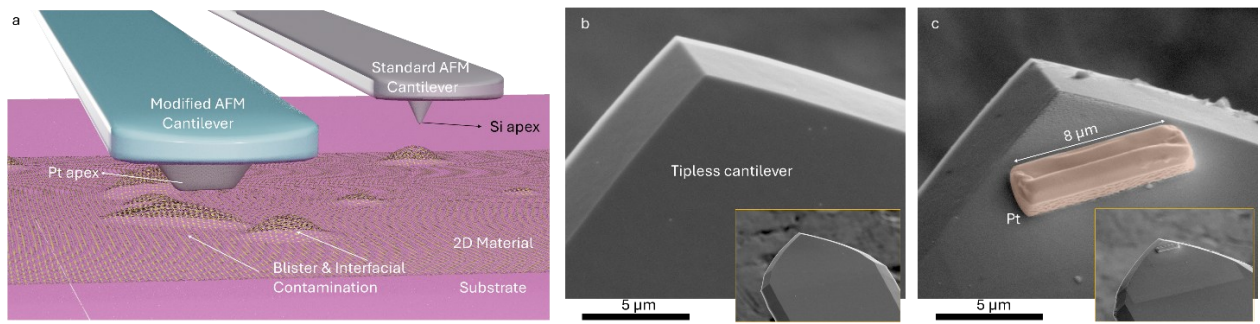

**Figure 1 a.** Schematic comparison of standard AFM cleaning vs. improved AFM cleaning. **b.** SEM image of a tipless AFM cantilever and **c.** after the Pt wedge apex is formed via FIB deposition.

Here, we introduce an improved AFM cleaning method, where, instead of using a pointy or spherical tip, we employ wedge-shaped tip geometries for cleaning 2D materials and heterostructures. The main advantage of our modified method is that minimal parameter optimization is required as compared to other AFM-based cleaning techniques. Due to the wide profile of our custom tip apex, higher throughputs are achieved as compared to pointy AFM tip cleaning, as fast as 3 $\mu m^2/s$. To showcase Pt-wedge squeegee cleaning, first, we demonstrated the improvement of photoluminescence linewidths of $WS_2$ monolayers on h-BN. Next, we showed how Pt-wedge can be used to control interfacial contaminants with Au electrodes and finally demonstrated that the method can be used to clean h-BN encapsulated heterostructures rapidly. A schematic comparison of our cantilevers to conventional cantilevers is shown in **Figure 1a**.

A wedge-shaped Pt apex is deposited on the underside of a tipless AFM cantilever, as shown in **Figure 1b-c**. We also demonstrated that a similar wedge-shaped Pt apex can be deposited on a conventional contact-mode AFM cantilever after clipping the tip apex with ion beam milling (**Figure S1**). AFM cantilever parameters are determined based on the force required to push a micrometer-sized blister of $MoS_2$ on $SiO_2$. When a blister is pushed to the side, the fluid inside the blister is pressurized at the leading edge to peel the sheet away from the surface[18]. The 2d sheet must bend and deform as the blister moves to the edges. The energy of this elastic deformation depends on the material and affects the delamination forces. Moreover, the adhesion energy, which is the energy required to separate the sheet from the surface per unit area of delamination, plays a crucial role in the entire process[19]. To simplify the calculation, we assume that the work done by the lateral force ($F$) over a small distance $dx$ is $F.dx$. This work must equal the energy required to create the new delaminated area[20], which is given by $G_A.2a.dx$. Here, the radius of the blister is denoted by $a$. As a result, $F \approx 2a.G_A$. Based on this equation, for a 1 μm wide blister of MoS$_2$ on SiO$_2$, using the adhesion energy reported in the literature[21] ~480 mJ m$^{-2}$, we find the force required to move the blister as ~2 μN. For the case of MoS$_2$ - h-BN heterostructure, the adhesion energy is reported[22] as ~136 mJ m$^{-2}$, which provides the required force to move the blister as ~0.5 μN. We assume MoS$_2$ and WS$_2$ would have similar adhesion energies.

Based on the estimation given above, we used a 450 μm long tipless silicon AFM cantilever with a spring constant of ~0.7 N/m. Then, to create a custom AFM tip, we deposited a Pt wedge on a tipless AFM cantilever using focused ion beam deposition. An 8 μm wide 300 nm tall wedge-shaped apex is deposited as shown in **Figure 1b-c**. This wedge creates a wide surface along the direction of the scan. The Pt wedge is more durable because the pressure is distributed over a larger contact area, and Pt offers an elastically deformable surface, unlike Si, which is more brittle. As another approach, we also used a conventional contact-mode AFM cantilever ($k \sim 4$ N/m), but we trimmed the cantilever apexes using FIB milling and deposited a Pt wedge on the trimmed apexes (**Figure S1**). Details regarding these stiffer cantilevers are provided in the Supporting Information. As mentioned in the subsequent text, some cleaning has been performed using the tipless cantilevers, and others using the clipped-tip cantilevers.

To demonstrate the effectiveness of the modified AFM tips in cleaning monolayer samples, we first performed AFM cleaning on a WS$_2$ monolayer that was partially placed on an h-BN flake and partially on the Si$_3$N$_4$-coated substrate. **Figure 2a** shows the optical micrograph of the sample. Before cleaning, a tapping-mode scan is performed with a standard AFM tip ($k \sim 1-5$ N/m), as shown in **Figure 2b**. The height trace map shows the blister formation, especially in the h-BN region. To assess how the photoluminescence (PL) spectrum is affected by the presence of the blisters, we performed PL mapping. **Figure 2c** shows the full-width-half-maximum (FWHM) value of the Lorentzian fit to the PL peak at each point. The intensity map is given in the Supporting Information.

Next, we performed a single-scan contact-mode cleaning with the Pt-deposited AFM cantilever at 0.8 V (102.5 nN). Details regarding the force calibration of the cantilevers are provided in the Supporting Information. A 30×30 μm² region is cleaned using a 256-line scan. This corresponds to 117 nm steps. Since the apex is 8 μm wide, by 4 μm from the start of the scan, each point is swept about 70 times. This provides ~3 μm²/sec. cleaning rate. This is an almost 3 orders of magnitude improvement over conventional AFM tip cleaning, which requires ~10 scans to achieve a similar level of cleanliness[23]. Although there is no significant change in the optical micrograph after the cleaning scan (**Figure 2d**), the tapping mode AFM height trace map, **Figure 2e**, shows the removal of blisters in both the h-BN and Si$_3$N$_4$-supported regions of WS$_2$. A more dramatic reduction in the blister height is shown in **Figure S2** in the supporting information. PL FWHM map shows narrowing after the cleaning scan (**Figure 2f**). Even in the flat regions of the sample, X$^0$ peak FWHM narrows as shown in **Figure 2g**. PL intensity maps are given in **Figure S3**. The effectiveness of the modified tip cleaning is evident from the tapping-mode scan over the cleaned and uncleaned regions (**Figure 2h**).

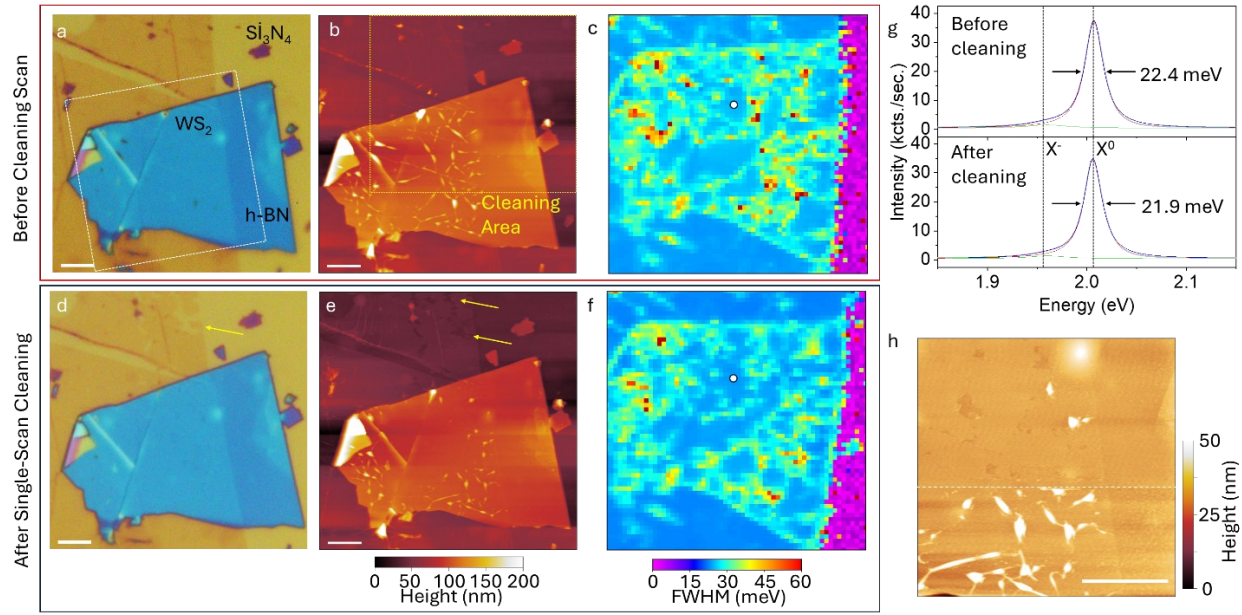

**Figure 2 a.** Optical micrograph of a $WS_2$ monolayer partially placed on h-BN on a $Si_3N_4$-coated substrate. The white dashed rectangle shows the region of PL mapping. **b.** The AFM height trace map collected in the tapping mode before AFM cleaning is shown. The cleaning area is marked. **c.** The full-width at half-maximum (FWHM) distribution of the PL signal is shown in the map before AFM cleaning. **d.** An optical micrograph of the sample after AFM cleaning is shown. The yellow arrow indicates the damaged region on $Si_3N_4$. **e.** The AFM height trace map after a single AFM cleaning scan shows the contrast between cleaned and uncleaned regions of the sample. Yellow arrows indicate regions of monolayer damaged by the cleaning process. **f.** FWHM map shows a consistent reduction in the cleaned region. **g.** PL intensity shows the narrowing of the neutral excitation FWHM after cleaning. White dots in **c** and **f** show the points where PL spectra are collected. **h.** A closer AFM scan of the boundary between the cleaned and uncleaned regions shows the effectiveness of the modified AFM tips for the cleaning. Scale bars are 5 μm.

Next, we focus on AFM cleaning of $WS_2$ on a 50/5 nm Au/Ti electrode patterned on a quartz substrate. As we have recently demonstrated, $WS_2$ on the Au surface can exhibit PL intensity enhancement due to screened exciton-exciton interactions[7]. This PL enhancement can be quenched by annealing the heterostructure. The van der Waals gap between $WS_2$ and Au collapses upon removal of the interfacial contaminants through annealing, allowing free charge carriers in the metal to be transferred to TMDC. To illustrate that AFM cleaning can remove interfacial contaminants, we first cleaned the gold surface with AFM to remove any large contaminants. Then the $WS_2$ monolayer is placed on the contact. First, we collected AFM height traces and PL intensity maps before the contact mode cleaning scan. Then, we repeated the same measurements after two cleaning passes, one at a set-point voltage of 0.2 V (27.2 nN) to confirm the cleaning region and the other at 0.7 V (89.7 nN), with a speed of 25 μm/s. The PL intensity has decreased by three times in the cleaned region. Apparent crystal thickness has decreased from 4.8 nm to 3.4 nm (**Figure S4**), indicating that the van der Waals gap between the monolayer and the Au has decreased.

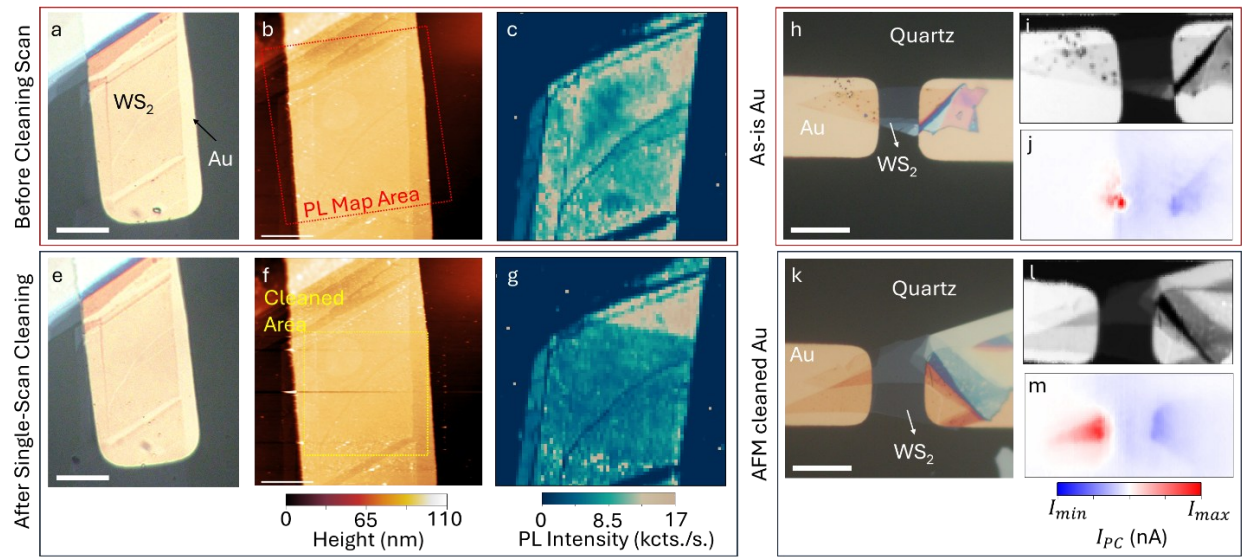

**Figure 3 a.** Optical microscope image of a WS$_2$ monolayer on an Au electrode. The scale bar represents 10 μm. **b.** AFM height trace map in tapping mode, and the red square marks the area of **c.** PL intensity map. Graphs through **a-c** are collected before AFM cleaning. **d.** Optical microscope image of the same sample, **e.** height trace map after cleaning, with the cleaned area marked by a yellow rectangle and **f.** PL intensity map. The map shows the contrast between cleaned and uncleaned areas. **h.** Optical microscope micrograph of WS$_2$ multilayer deposited on gold electrodes without any AFM cleaning on the Au surface. SPCM **i.** reflection map and **j.** photocurrent map of the sample is shown. The photocurrent map shows a speckle-like photocurrent distribution over the contacts, indicating poor electrical junction formation. **k.** Optical microscope micrograph of WS$_2$ multilayer deposited on AFM cleaned gold electrodes. SPCM **l.** reflection map and **m.** photocurrent map of the sample is shown. A uniform photocurrent distribution is evident on cleaned electrodes, indicating better electrical junction formation.

The increase in cleaning rate and the durability of the tips enabled practical cleaning of the substrate and pre-patterned metal contacts. As a demonstration, we prepared two-terminal gold devices on a quartz substrate without AFM cleaning of electrodes (**Figure 3h-j**) and with AFM cleaning of electrodes (**Figure 3k-m**). A single cleaning scan of the contact electrode surface is followed by the dry transfer of a WS$_2$ flake. Then, we performed scanning photocurrent microscopy (SPCM) on both samples. SPCM here acts as a spatially resolved probe of the contact resistance map. WS$_2$ generates a photothermoelectric-dominated response at the contacts[24]. When the contact resistance is large, typically, local photoresponse from the points where the crystal makes contact with the electrode is observed[25]. However, high-quality contacts often reveal a uniformly distributed photoresponse from the entire crystal area. This has been demonstrated in annealed samples before[26]. **Figure 3h-m** shows our measurements from cleaned and uncleaned contacts and exhibits that a single cleaning of the gold electrode surface with the modified cantilever effectively removes the surface contaminants. This approach can be particularly important in applications that are sensitive to high-temperature annealing, such as low thermal budget back-end of the line integration of two-dimensional materials[27].

Fabricating high-quality interfaces for heterostructures of atomically thin materials encapsulated with h-BN is another important application of AFM-based cleaning[28,29]. Our improved method is also applicable to such heterostructures as showcased in **Figure 4a-d.** We fabricated a WS$_2$/MoS$_2$ stack on h-BN, encapsulated by another h-BN layer. All 2D materials are first exfoliated on an elastomer and deterministically transferred on top of each other using a micromanipulator. No cleaning or annealing is performed between the transfer stages to exacerbate contamination. Both the optical microscope image (**Figure *4*a**) and the AFM tapping mode height trace scan (**Figure *4*b**) show the presence of liquid-like contaminants across the flakes. After AFM cleaning with the clipped-tip cantilever at 1.84 µN contact force, interfacial contaminants are dragged to the edges of the scan area, and a noticeable reduction in wrinkles is apparent (**Figure *4*c-d**). Larger forces are used in this cleaning run as the h-BN top layer is significantly stiffer than the monolayers on h-BN. Further cleaning scans would improve cleaning performance.

As Pt is an elastically deformable metal and the contact forces are limited due to the large contact area, we observe that the Pt apex is very durable. Even after more than 100 cleaning scans, the tip retained its structure, with only contaminants from the cleaning scans attached to its tip. A comparison between right after Pt wedge deposition and after more than 100 cleaning scans is provided in **Figure *4*e-f**. However, we would like to note that Pt may not be the best choice of metal for cleaning hydrocarbons, especially for removing adventitious carbon contamination on the surface of 2D materials. Pt has very low reactivity toward hydrocarbon chains[30]. We propose that a metal that can form carbides, such as Cr, Ti, or W, would be a better choice for improved cleaning performance. However, controlled deposition of these metals as a wedge at the tip of an AFM cantilever is more challenging than that of Pt. We would like to leave this proposition as an open challenge for future studies.

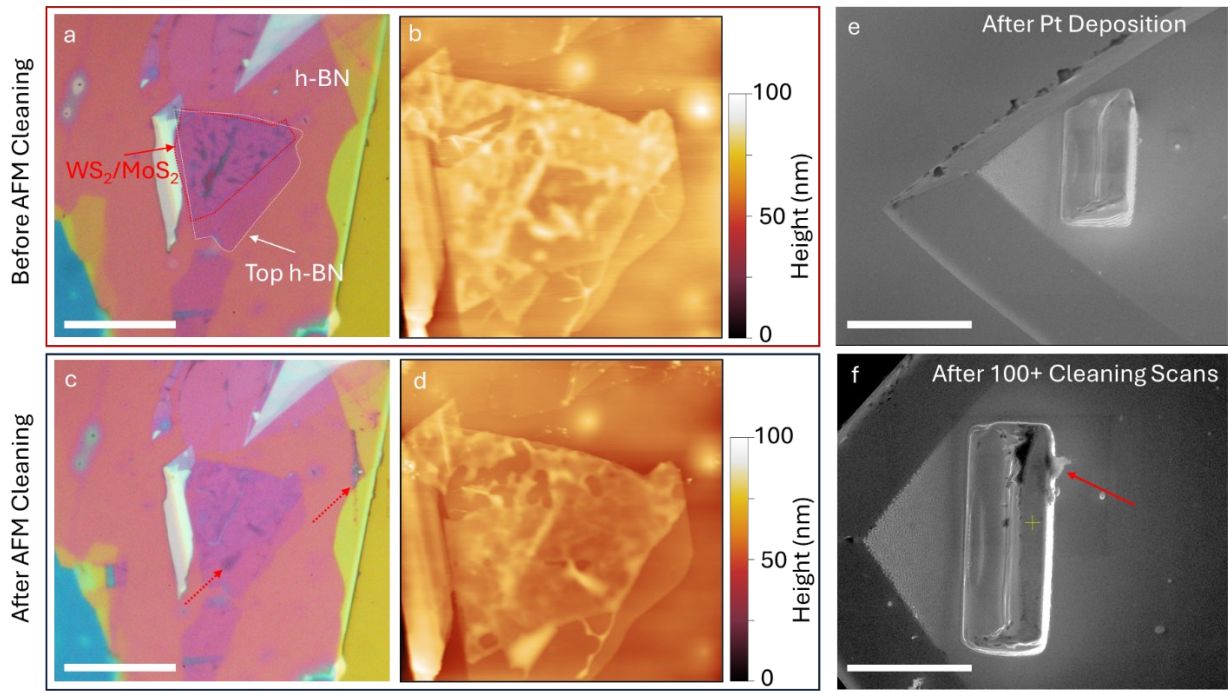

**Figure 4 a.** Optical microscope micrograph of an h-BN/WS$_2$/MoS$_2$/h-BN stack on a Si$_3$N$_4$ substrate. Scale bar represents 20 µm. **b.** Before cleaning AFM height trace map of the h-BN encapsulated area of the stack. **c.** Optical microscope image of the stack in **a** after the cleaning scan. **d.** AFM height trace map after the cleaning scan. **e.** SEM image of Pt wedge after fabrication. **f.** SEM image of the same cantilever after more than 100 contact mode cleaning scans. Scale bars are 5 µm.

In conclusion, we have introduced a modified AFM cleaning method that speeds up the cleaning process by four orders of magnitude, thanks to the much larger surface area provided by the wedge-shaped AFM apex.

This method not only decreases the cleaning time but also simplifies the procedure compared to using conventional AFM tips. Moreover, we demonstrated that conventional contact-mode tips can be modified by first clipping the apex with ion-beam milling, followed by Pt wedge deposition, thereby improving the method's versatility. To showcase the method, we demonstrated its applicability for cleaning various 2D heterostructures and for integrating them with 3D electrodes. The method can also be used in twisted heterostructures. Finally, we showed that a Pt tip can be used more than 100 times without a significant change in its cleaning performance. We reckon our method will improve the widespread applicability of AFM-based cleaning of two-dimensional materials and their interfaces, especially if a commercial wedge-cleaning tip becomes available.

## Acknowledgements

TSK acknowledges support by the Turkish Scientific and Technological Research Council under grant no 125F207.

## Author Contributions

EY performed the AFM-related experiments and optical and electrical characterizations. DP prepared the modified AFM cantilevers and metallic patterns. TSK conceived the idea, supervised the project, and wrote the manuscript with contributions from EY and DP.

# Supporting Information: Pt-wedge squeegee cleaning of two-dimensional materials and heterostructures

Emine Yegin[1,2], Doruk Pehlivanoğlu[1,2], T. Serkan Kasırga[2,3*]

[1] Institute of Materials Science and Nanotechnology, Bilkent University, Ankara 06800 Türkiye
[2] Bilkent National Nanotechnology Research Center – UNAM, Ankara 06800 Türkiye
[3] Department of Physics Middle East Technical University, Ankara 06800 Türkiye
[*] Corresponding author: kasirga@bilkent.edu.tr

## 1. Experimental Methods

**Sample exfoliation and transfer.** Multilayer hBN flakes were mechanically exfoliated from bulk crystals and transferred onto a clean $Si_3N_4$/Si substrate using a PDMS-assisted dry-transfer technique. The flakes supported on a PDMS stamp (Gel-Pak) were first located under an optical microscope, then carefully aligned with the target substrate using a micromanipulator, and finally released during the stamping process. Subsequently, a monolayer $WS_2$ flake was exfoliated and transferred onto the hBN layer. By repeating the same procedure, the complete hBN/$WS_2$/$MoS_2$/hBN heterostructure and few-layer $WS_2$ flakes were also transferred onto $Si_3N_4$/Si substrates and thermally evaporated 50/5 nm Au/Ti contacts, respectively.

**Atomic Force Microscopy (AFM) measurements.** AFM measurements were performed using an Asylum Research MFP-3D. For topography imaging, cantilevers with spring constants of $k \approx 1–5$ N/m were used, while for contact mode measurements, both $k \approx 1–5$ N/m and $k \approx 0.2–0.7$ N/m (tipless) cantilevers were employed from Bruker and NanoSensors, respectively. Each AFM tip was calibrated using the thermal method to determine the inverse optical lever sensitivity (InvOLS) and the spring constant. The tip force in contact mode was calculated using the relation: deflection * spring constant * InvOLS. AFM probe maintained a constant force during both trace and retrace scans. Tapping mode was used for topography imaging, while contact mode was applied to remove interfacial contaminants and bubbles. Scan speeds ranged from 20 μm/s to 30 μm/s, with a resolution of 256 × 256 pixels.

**Scanning Photocurrent Microscopy (SPCM) measurements.** For SPCM measurements, a commercial setup from LST Scientific Instruments Ltd. was used. The system has a small scanning head with a laser that is simple to swap out. One SR830 lock-in amplifier was used to record the reflection map, while the other was utilized to measure photocurrent. A 40× objective lens was used to focus a 633 nm laser onto the sample, and the scan was performed using a 500 nm step size at zero bias. Laser power measured 95 μW by Thorlabs. A lock-in amplifier was used to test the electrical response of the devices using tungsten probes positioned on the gold contact pads. This signal was then compared to the laser's mechanical modulation frequency, which is roughly 800 Hz.

**Photoluminescence (PL) measurements.** A WITec Alpha 300S system with a 532 nm excitation laser and a spectrometer connected to a cooled CCD detector was used to perform PL measurements. After the objective, a Thorlabs power meter was used to measure the laser power, which was 3 μW for mapping. PL maps covering an area of 37 × 37 μm² were obtained with a resolution of 70 × 70 pixels (points per line × lines per image) and a scan time of 70 s/line.

## 2. Fabrication of a modified apex on a standard contact-mode AFM cantilever

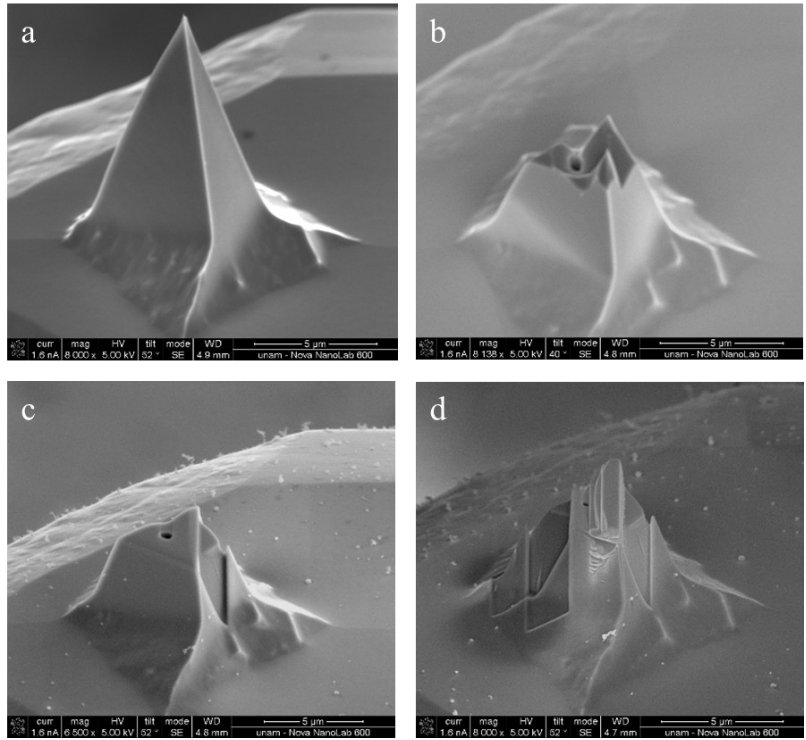

**Figure S1** SEM images of pristine and modified contact mode AFM tip (k = 1-5 N/m). **a.** Pristine tip, and **b.** damaged tip from the same batch. **c.** FIB milled apex for Pt wedge deposition preparation. **d.** After the Pt wedge deposition.

## 3. Cleaning of WS$_2$ on h-BN

**Figure S2** shows a WS$_2$ monolayer transferred onto h-BN on a Si$_3$N$_4$ substrate. The cantilever shown in **Figure S1** is employed for the cleaning. 0.6 µN force is applied (Set point voltage: 0.45 V) with a scan speed of 25 µm/s. The scan area is 60x30 µm$^2$, with a total scan time of ~600 seconds (trace and retrace), at a rate of ~3 µm$^2$/s.

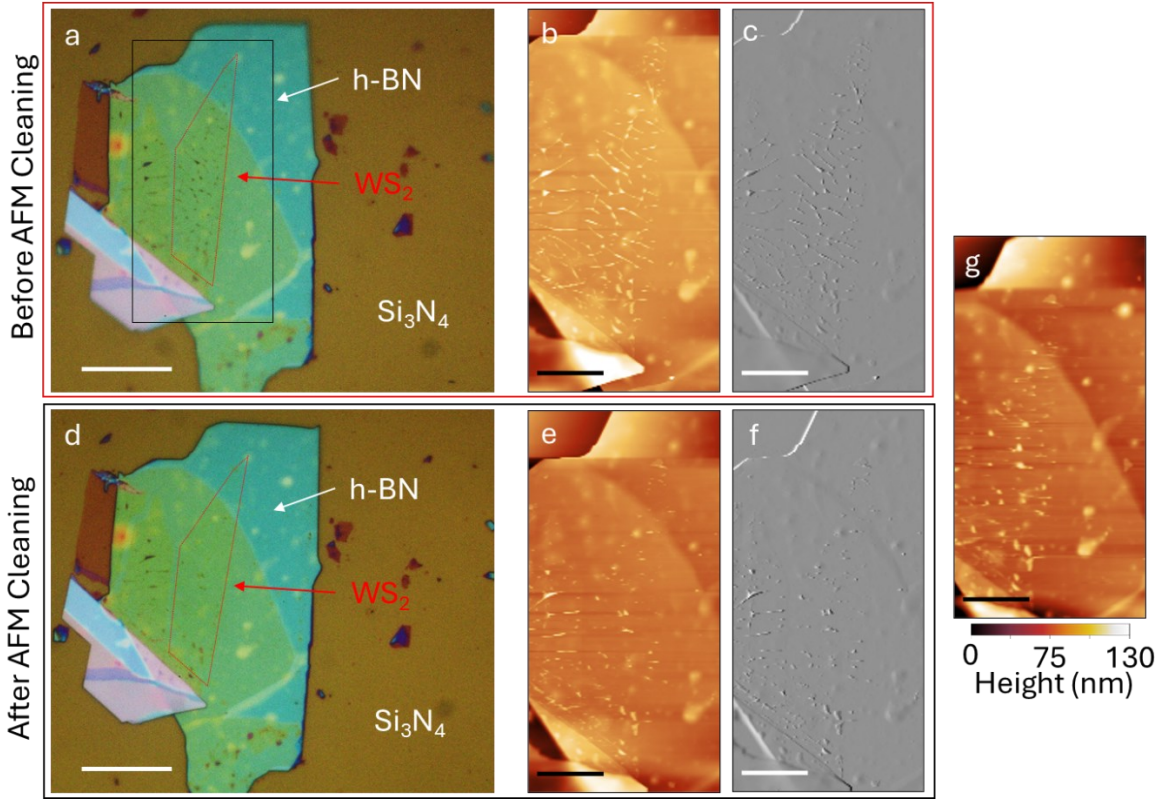

**Figure S2 a.** Optical microscope image of a WS$_2$ monolayer on h-BN stacked on a Si$_3$N$_4$ substrate is shown. The black rectangle shows the AFM scan and cleaning area. Scale bar is 20 μm. **b.** AFM height trace map before AFM cleaning. **c.** Amplitude retrace of the AFM scan is plotted to show the contrast of interfacial contaminants before cleaning. **d.** Optical microscope image after AFM cleaning. **e.** AFM height trace map after AFM cleaning. Scale bar is 20 μm. **f.** Amplitude retrace shows a significant difference as compared to the before-cleaning scan. **g.** The height trace map collected during the contact-mode cleaning scan is provided for reference. Despite the large contact area of the cantilever, a decent resolution image can be obtained during the scan for reference. All height trace maps share the same color bar. Scale bars for AFM scans are 5 μm.

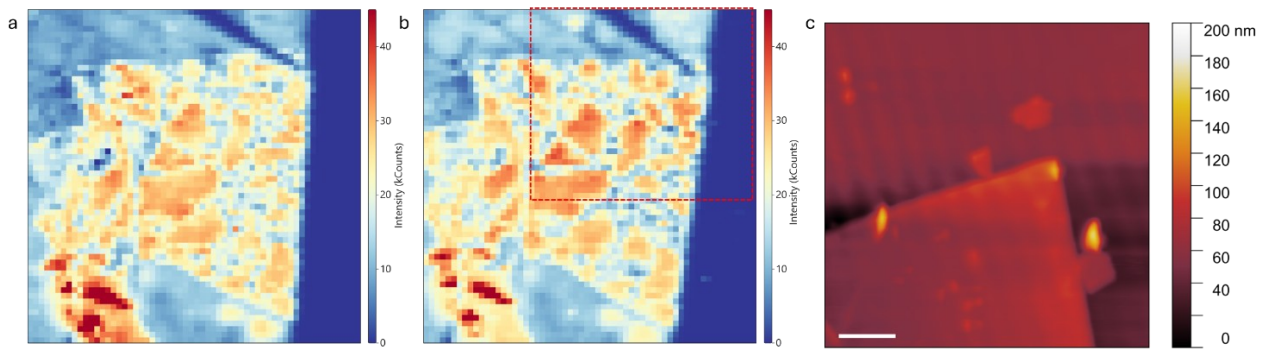

**Figure S3 a.** PL intensity map of the sample shown in Figure 2 in the main text before the cleaning scan. **b.** PL intensity map of the sample after the cleaning scan. The red-dashed rectangle shows the cleaned Pt-wedge area. Scan size is 25x25 μm$^2$. **c.** Height trace map collected during the cleaning scan at 103 nN. Scale bar is 5 μm.

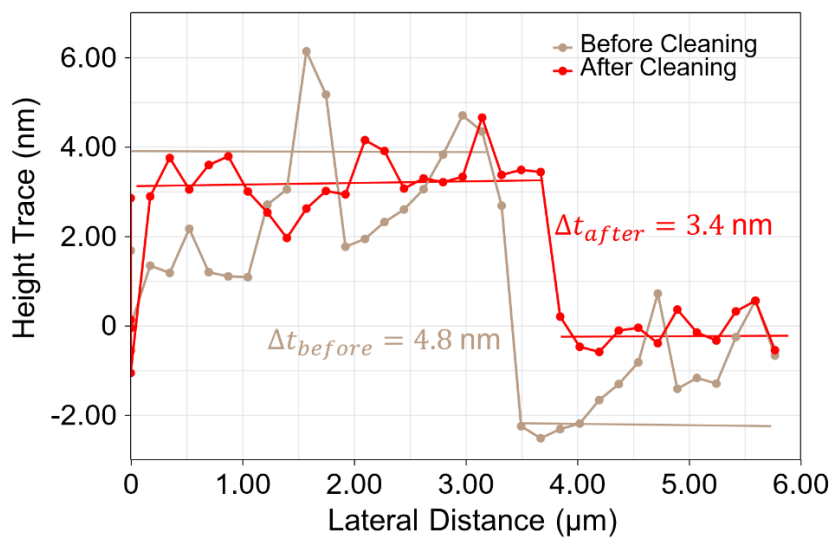

**Figure S4** AFM height traces taken from cleaned and uncleaned regions of the crystal.

### 4. Contact mode cleaning force calibration

To calibrate the force applied to the sample by the modified AFM cantilevers, first, inverse optical lever sensitivity (InvOLS) is measured against a $Si_3N_4$-coated Si substrate. The slope of the force-distance curve provides the InvOLS parameter for the tips. Then, using the thermal noise method, we extract the spring constant of the cantilever and confirm that it falls within the producer's specifications. Finally, for each set voltage, we calculate the force applied by the cantilever to the surface. For the tipless cantilever with Pt deposited at the apex, we found InvOLS to be 476 ± 5 nm/V and the spring constant to be 0.275±0.010 N/m. For the trimmed cantilever, InvOLS is 330 ± 2 nm/V, and the spring constant is 4.05 ± 0.05 N/m.